\documentclass[prl,reprint]{revtex4-1}

\draft 

\usepackage{color}
\usepackage{graphicx}
\usepackage{units}
\usepackage{amssymb}
\usepackage{lineno}
\usepackage{footmisc}
\usepackage{perpage}

\begin{document}




\makeatletter
\def\blfootnote{\xdef\@thefnmark{}\@footnotetext}
\makeatother

\interfootnotelinepenalty=10000

\begin{widetext}
\thispagestyle{empty}

\oddsidemargin=-2cm
\textwidth 167mm \textheight 245mm \voffset 16mm \hoffset 17mm
\baselineskip18pt\parskip5pt\parindent17pt

\vspace*{-1.4cm}
\textbf{FOREWORD} \\


The first attempt to apply a mathematical framework to thermocapillary flows is attributed to the article written in 1959 by N.O. Young, J.S. Goldstein and M.J. Block [YGB]. This classic work brought interface driven flows into attention of the Eastern hydrodynamic community particularly focusing on the surface flows induced by the temperature inhomogeneities.
However, there is another work, which is currently hidden from worldwide scientific attention due to the fact that it was written in Russian many years ago. In 1956, in Zhurnal Fizicheskoi Khimii (a journal of the former USSR, nowadays it is known
as Russian Journal of Physical Chemistry A) A.I. Fedosov published an article entitled `Thermocapillary motion', where he introduced a mathematical description of the thermocapillary effect, sequentially considering two problems: the motion of a flat liquid layer and the motion of a spherical non-deformed drop without gravity.
After thorough historical research, and with much help from different people we found that this result was obtained by Fedosov before the year 1948, in his doctorate thesis (under the supervision of Benjamin Levich). However, the story is even more curious: in 1944, Lev Landau and Evgeny Lifshitz published [LL] the most general form of the boundary conditions for the liquid-liquid interfaces which ultimately leads to any cause of surface-driven motion, including the thermocapillary one.
Below, we present an English translation of the Fedosov article, temporarily leaving aside the chronology of the scientific inputs of Fedosov, Levich and Landau, and reserving to return to this story in the near future.

In a course of this translation, a substantial amount of historical work was done. We deeply appreciate Sergey A.~Fedosov, Alexey V.~Belyaev, and Igor Yu.~ Makarikhin for their generous help in obtaining the important documents. We are also indebted to Michael  K\"{o}pf for critical reading of this translation.

[YGB] N.O. Young,  J.S. Goldstein, and M. J. Block, \emph{The Motion of Bubbles
in a Vertical Temperature Gradient}, J. Fluid Mech. \textbf{6}, 350-356 (1959).

[LL] L.D. Landau and E.M. Lifshitz, \emph{Mechanics of Continuous Media} (OGIZ, Moscow, 1944) (in Russian).

\vspace{1cm}

March, 2013, Canada, Israel \hspace{4.9 cm} V.~Berejnov,\,\, K.I.~Morozov

\vspace{2cm}

Dr. Viatcheslav Berejnov  \hspace{5.5 cm} Dr. Konstantin I. Morozov
\vspace{-0.4cm}

Brockhouse Institute for Materials Research \hspace{2.63 cm} Department of Chemical Engineering
\vspace{-0.4cm}

McMaster University   \hspace{6.17 cm} Technion -- Israel Institute of Technology
\vspace{-0.4cm}

Hamilton, Canada \hspace{6.61 cm} Haifa 32000, Israel
\vspace{-0.4cm}

\underline{berejnov@gmail.com} \hspace{6.31 cm}  \underline{mrk@tx.technion.ac.il}

\clearpage

\end{widetext}

\pagestyle{plain}

\setcounter{page}{1}


\textmd{\underline{\scriptsize{Zhurnal Fizicheskoi Khimii 30, N2, p. 366-373 (1956)}}}\\ \\ \\

\title{\large{THERMOCAPILLARY MOTION}}

\author{A. I. Fedosov}


\affiliation{Chita State Pedagogical Institute, Russia}

\author{(Received 16 May 1955; published February 1956)}


\begin{abstract}

Mechanical motion (convection) appears in a nonuniformly heated liquid because its density depends on temperature$^{1)}$. This flow causes the liquid to self-mix unifying the temperature field. The convective flows are addressed in a series of papers recently published [1].
When the liquid has an interface, besides the convection, a different kind of a motion can appear. The origin of this motion is a gradient of the surface tension$^{2)}$. Similarly to the electrocapillary motion, one can call this new motion a thermocapillary motion. In the present paper we consider two classes of thermocapillary motion: motion of a liquid in an open container and motion of a liquid drop suspended in another liquid.

\end{abstract}

\maketitle

\subsection{Motion of liquid in a flat open container when a temperature gradient is applied along the liquid interface}

Considering the motion of liquid in a flat container let us assume that width and length of the container are sufficiently larger than its depth.
This assumption
reduces considerably the complexity of equations of liquid motion.
The component momentum equations for steady motion of a viscous incompressible liquid written in the Cartesian coordinates are
\begin{eqnarray}
v_x \frac{\partial v_x}{\partial x}+v_y \frac{\partial v_x}{\partial y}+v_z \frac{\partial v_x}{\partial z}=-\frac {1}{\rho}\frac{\partial p}{\partial x}+\nu \Delta v_x+F_x,\nonumber
\end{eqnarray}
\vspace*{-0.4cm}
\begin{eqnarray}
v_x \frac{\partial v_y}{\partial x}+v_y \frac{\partial v_y}{\partial y}+v_z \frac{\partial v_y}{\partial z}=-\frac {1}{\rho}\frac{\partial p}{\partial y}+\nu \Delta v_y+F_y\,,
\label{eq:1}
\end{eqnarray}
\vspace*{-0.4cm}
\begin{eqnarray}
v_x \frac{\partial v_z}{\partial x}+v_y \frac{\partial v_z}{\partial y}+v_z \frac{\partial v_z}{\partial z}=-\frac {1}{\rho}\frac{\partial p}{\partial z}+\nu \Delta v_z+F_z\,.\nonumber
\end{eqnarray}
Here $\textbf{\emph{F}}$  is an external body force$^{3)}$\blfootnote{$^{1)}$ This is true with gravity applied (KM \& SB).
\\$^{2)}$ Currently, this kind of motion is known as Marangoni convection in contrast to gravitational Rayleigh-B\'enard convection (KM \& SB). \\ $^{3)}$ $\nu=\mu/\rho$, where $\nu$, $\mu$, and $\rho$ are kinematic viscosity,
dynamic viscosity, and density of the liquid, respectively (KM \& SB).}.

Let us introduce an origin of coordinate system on the interface of liquid. The $x$- and $y$-axes are applied along the container length and width, respectively, the $z$-axis is directed from the liquid interface down to the bottom of the container.
We assume a constant temperature gradient is applied along
the $x$-axis, giving us $v_y=0$.


We will also neglect the convective flow because the depth of the container is smaller than its length.
Feasibility of this assumption and the limitations which it causes for the theory will be addressed below.
Now, since we disregarded the convective motion, then far away from the container walls $v_z=0$. Further, the term $\partial v_x/\partial x \sim v_x/l$ ($l$ is the container length) is much smaller than $\partial v_x/\partial z \sim v_x/h$ ($h$ is a depth of the container) and can be
omitted, too. As a result of this simplifications, the system (1) is reduced to only two equations. The first one is
\begin{equation}
 \frac{\partial p}{\partial x}=\mu \frac {\partial ^2 v_x}{\partial z^2}\,.
\vspace*{0.2cm}
\label{eq:2}
\end{equation}
Liquid in the container is dragged by the moving surface layer, then it returns backward due to the container walls causing the appearance of a pressure drop along the $x$-axis.
The second equation is
\begin{equation}
 \frac{\partial p}{\partial x}=\rho g\,.
\label{eq:3}
\end{equation}
After integrating this equation the solution is
\begin{equation}
p=p_1(x)+\rho gz\,.
\label{eq:4}
\end{equation}
Now we will define the boundary conditions for Eq.~(2). First, on a solid boundary the liquid velocity vanishes,
it gives us a first condition
\begin{equation}
v_x(z=h)=0\,.
\label{eq:5}
\end{equation}
Second, on a free liquid interface the components of a viscous-stress tensor are continuous.
Since $v_y=v_z=0$, then
only the one tensor component $p_{xz}$ is essential for us while
the others are either equal to zero or are not important at all.
The continuity of $p_{xz}$ results to the second
boundary condition
\begin{equation}
\frac {\partial \sigma}{\partial x}=-{\mu \frac {\partial v_x}{\partial z}}\,\,\,\,\,\textrm{at}\,\,\,{z=0}\,,
\label{eq:6}
\end{equation}
where $\sigma$ is the surface tension.
Besides the boundary conditions, the solution of Eq.~(2) should meet an additional condition
representing the fact that the average velocity over the container cross section equals to zero:
\begin{equation}
\frac {1}{h}\int_0^h v_x dz=0\,.
\label{eq:7}
\end{equation}
With due account of relation (4), equation (2) is easily integrated. Indeed, substituting  Eq.~(4) in Eq.~(2) we have
\begin{eqnarray}
 \frac{d p_1(x)}{d x}=\mu \frac {d ^2 v_x}{d z^2}\,.
\nonumber
\end{eqnarray}
Since $p_1(x)$ does not depend on the coordinate $z$, the solution of Eq.~(2) is
\begin{eqnarray}
v_x=A+Bz+\frac {1}{2\mu} \frac {d p_1(x)}{d x}z^2\,.
\nonumber
\end{eqnarray}
%

Applying the boundary conditions we have
\begin{eqnarray}
B=-\frac {1}{\mu} \frac {d\sigma}{dx}\,,
\nonumber
\end{eqnarray}
\begin{eqnarray}
A=\frac {h}{\mu} \frac {d\sigma}{dx}-\frac {1}{2\mu} \frac {d p_1(x)}{d x}h^2\,,
\nonumber
\end{eqnarray}
and finally the expression for the velocity is
\begin{eqnarray}
v_x=\frac {1}{\mu} \frac {d\sigma}{dx}(h-z)-\frac {1}{2\mu} \frac {d p_1(x)}{d x}(h^2-z^2)\,.
\nonumber
\end{eqnarray}
Applying this formula to condition (7) gives
\begin{eqnarray}
\frac {dp_1(x)}{dx}=\frac {3}{2h} \frac {d\sigma}{dx}\,,
\nonumber
\end{eqnarray}
and for the pressure in the liquid we have
\begin{equation}
p=p_0+\frac {3}{2h} [\sigma(x)-\sigma(0)]+\rho g z\,,
\label{eq:8}
\end{equation}
where the constant $p_0$ cannot be found
because the pressure in liquid is always defined to the extent of an additive constant.
Finally, it gives us the velocity function in the liquid
\begin{equation}
v_x=\frac {1}{4\mu h}\frac {d\sigma}{dx} (3z^2-4hz+h^2)\,,
\label{eq:9}
\end{equation}
or
\begin{equation}
v_x=\frac {1}{4\mu h}\frac {d\sigma}{dT} \frac {dT}{dx}(3z^2-4hz+h^2)\,.
\label{eq:10}
\end{equation}
The profile of the velocity field is shown in Fig.~1.
\begin{figure}[t] \centering
\includegraphics[width=5 cm]{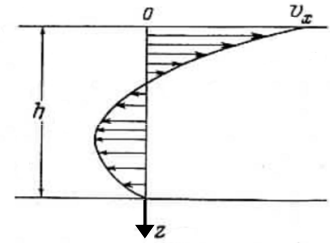}
\caption{
The profile of the velocity field for thermocapillary motion in a flat long container.}
\end{figure}

One can see from Eq.~(10) that velocity of the liquid reaches its maximum on the liquid interface
\begin{equation}
v_{x,\,max}=\frac {h}{4\mu }\frac {d\sigma}{dT} \frac {dT}{dx}\,.
\label{eq:11}
\end{equation}
Since ${d\sigma}/{dT}<0$, therefore along the interface, the velocity direction is
opposite to the temperature gradient.



\vspace{0.5cm}

If one compares the expression for the maximal velocity of the thermocapillary flow$^{4)}$\blfootnote{$^{4)}$ see Eq.~(11) (KM \& SB).} against the expression for the characteristic velocity of the convective flow, introduced below, then one may think that the convective flow could be simply disregarded for the condition of large container depth $h$. However, it is not the case. In fact, formula (11)
cannot be used for any size of the container.
Indeed, in the momentum equations we have neglected the term
$v_x ({\partial v_x}/{\partial x})\sim {v_x^2}/{l}$ comparing to
$\nu ({\partial^2 v_x}/{\partial z^2})\sim \nu {| v_x |}/{h^2}$.
In other words, we assumed that ${v_x^2}/{l} \ll \nu {| v_x |}/{h^2}$ or,
equally, $h^2 \ll  {\nu l}/{| v_x |}$.
Substituting Eq.~(11) into the last expression, we have a validity criterion for the developed theory:
\begin{equation}
h^3 \ll \frac {4\nu ^2 \rho l}  {\left|\frac {d\sigma}{dT} \frac {dT}{dx} \right|}\,.
\label{eq:12}
\end{equation}
When this inequality is not valid, our theory is not valid too.
Thus, we cannot argue that if the depth, $h$,  of the container is large enough, then velocity of the thermocapillary flow overcomes velocity of the convective flow. One would also think that with unlimited growth of
both sizes $h$ and $l$, the thermocapillary velocity could exceed its convective
value, however it is not the case as well.
There are two restrictions regarding this way.

a) The temperature gradient is constant for any size of the container.
In this case, the temperature difference $\Delta T$ over the container is $\Delta T=T-T_0=\nabla T l\sim h^3$, and the convective velocity$^{5)}$\blfootnote{$^{5)}$ see Eq.~(*) below (KM \& SB).} grows much
faster than the thermocapillary one because the former depends on
temperature difference $\Delta T$ and not on temperature gradient.

b) The temperature difference is constant.
The expression for the maximal thermocapillary velocity can be rewritten as
\begin{eqnarray}
| v_{x, \,max} |= \frac {1}{\mu} {\left|\frac {d\sigma}{dT}\right|} \frac {|\Delta T |}{l}h\,.
\nonumber
\end{eqnarray}
Consider the estimate most favorable for the thermocapillary flow$^{6)}$\blfootnote{$^{6)}$ a limit of validity for the theory
(KM \& SB).}
\begin{eqnarray}
h^3 \sim \frac {\nu ^2 \rho l}  {\frac {|\Delta T|}{l}\left|\frac {d\sigma}{dT}\right|}\,.
\nonumber
\end{eqnarray}
Then we have
\begin{eqnarray}
| v_{x, \,max} |\sim  \displaystyle{\frac {|\Delta T|^{{2/3}}\left|\frac {d\sigma}{dT}\right|^{2/3}}{\nu^{1/3}\rho^{2/3}l^{1/3}}} \,.
\nonumber
\end{eqnarray}
Thus, the maximal velocity increases with growth of the temperature difference and temperature coefficient of
surface tension and decreases with the growth of viscosity and density. This is obvious.
The most important result is the dependence of maximal velocity on the container size $l$.
It is seen that under fixed temperature difference the maximal velocity  of thermocapillary flow decreases with growth of $l$.


Assuming the following values for the parameters $\mu=0.01$ P, $dT/dx=0.1$ K/cm,
$h=3$ cm, $d\sigma/dT=-0.15$ erg/(cm$^2$K), involved in the thermocapillary velocity expression, we calculate the maximal velocity value
$v_{x, \,max} =1.1$ cm/s $^{7)}$\blfootnote{$^{7)}$ The aforementioned part of the article is identical to the text in the PhD thesis [3] of Fedosov and to the part of the chapter `Thermocapillary motion' of the book [4]. Fedosov's next consideration
(the text in between two markers $\blacktriangledown$ and $\blacktriangle$)
of the convective contribution is not applicable to the given problem.  As it was shown by Birikh [5], for a thin layer the convective contribution can be found analytically in the same way as the thermocapillary one. The later prevails for layer with thickness $<1\div2$ mm, see details in [5] (KM \& SB).}.

\scriptsize

\vspace{0.5cm}

$\blacktriangledown$

\normalsize

Let us compare this estimate with the value of the convective velocity at the same conditions. The expression for
the convective velocity has a form [1] $^{8)}$\blfootnote{$^{8)}$ The problem considered in Ref.~[1] and cited here is the problem of a convective flow near the hot vertical wall. The geometry of the vertical wall problem is essentially different from the geometry of the horizontal layer studied here.
This mismatch between the geometries of the vertical walls and horizontal layer results to the incorrect estimate of the convective flow $J_{\mathrm{conv}}$ (see the next footnote) (KM \& SB).}
\begin{eqnarray}
\hspace{2.3cm}v_{\mathrm{conv}}=u_z\frac {x}{\delta}\left(1-\frac{x}{\delta}\right) ^2\,,\hspace{2.3cm}(*)
\nonumber
\end{eqnarray}
where $\delta$ is the boundary layer thickness, $x\le \delta$ and
\begin{eqnarray}
u_z=5.17\nu \left( \frac {\nu}{\gamma}+\frac {20}{21} \right)^{-1/2}\left[ \frac {g(T_1-T_0)}{\nu^2T_0}\right]^{1/2}z^{1/2}\,,
\nonumber
\end{eqnarray}
%

%
\begin{eqnarray}
\delta=3.93 \left( \frac {\gamma}{\nu}\right)^{1/2} \left( \frac {\nu}{\gamma}+\frac {20}{21} \right)^{1/4}\left[ \frac {g(T_1-T_0)}{\nu^2T_0}\right]^{-1/4}z^{1/4}\,.
\nonumber
\end{eqnarray}
Here $g$ is the gravity, $T_1$ and $T_0$ are the temperatures of the wall and far from the wall, respectively, $\nu$ is the kinematic viscosity,  $\gamma$ is the thermal diffusivity, $z$ is the nontrivial coordinate of a flat boundary layer.
Since Ref.~[1] does not provide an explicit expression for the  convective velocity outside the boundary layer, we have to compare liquid fluxes rather than velocities:
\begin{eqnarray}
J_{\mathrm{ conv}} =\rho \int_0^{\delta}u_{\mathrm{ conv}}dx=\frac {1}{6}u_z\delta \rho\,.
\nonumber
\end{eqnarray}
Substituting here $u_z$ and $\delta$ we find$^{9)}$\blfootnote{$^{9)}$ The relevant consideration of the convective term has been done in [5]. There was found that the convective flux (or velocity) $J_{\mathrm{ conv}}\sim \mathrm{Gr} \sim h^4$ ($\mathrm{Gr}$ is Grashof number), whereas the thermocapillary flux $J_{\mathrm{th-cap}}\sim \mathrm{Mg} \sim h^2$ ($\mathrm{Mg}$ is Marangoni number). In the case of sufficiently thin layer, the thermocapillary flow always prevails (see details in [5]). Fedosov's estimate proves to be a fourth root of the correct one, $J_{\mathrm{conv}}\sim \mathrm{Gr}^{1/4}$
(KM \& SB).}
\begin{eqnarray}
J_{\mathrm{ conv}} =3.4\mu\left( \frac {\gamma}{\nu}\right)^{1/2}  \left( \frac {\nu}{\gamma}+\frac {20}{21} \right)^{-1/4}\left[ \frac {g(T_1-T_0)}{\nu^2T_0}\right]^{1/4}h^{3/4}\,.
\nonumber
\end{eqnarray}
%
The corresponding relation for the thermocapillary flux equals
\begin{eqnarray}
J_{\mathrm{th-cap}} =\rho\int_0^{h/3}v_xdz=\frac {h^2}{27\nu}\frac {d\sigma}{dT}\frac{dT}{dx}\,.
\nonumber
\end{eqnarray}
Using the values of parameters mentioned above and $\rho=1$ g/cm$^3$, $T_1-T_0=1$ K, $T_0=300$ K, we obtain
the following
estimates for both flows: $ J_{\mathrm{ conv}} = 0.23$ g/(cm$\cdot$s) and $ J_{\mathrm{th-cap}} = 0.5$ g/(cm$\cdot$s).
Thus, the thermocapillary mass flux proves to be prevailing over the convective one.
Note, the given estimates are true for fairly large temperature gradients$^{10)}$\blfootnote{$^{10)}$ $dT/dx=0.1$ K/cm (KM \& SB).}.
If the temperature gradient was of the order of 0.01 K/cm $^{11)}$\blfootnote{$^{11)}$ e.g., more  realistic for the long and shallow containers than the one above (KM \& SB).}, our conclusion would be changed to the opposite one and now the convective flow would dominate.
Thus, one can conclude that the thermocapillary motion is negligible in all practically interesting situations.
However, this is not true for some experiments where the heating of liquid is caused by applying the light on its surface.
In this case a liquid is heated in a very thin layer near the free surface.
For the considered termocapillary problem, this `skin layer' doest not mean anything because
the thermocapillary velocity depends on the total thickness of the liquid layer.
For the convective flow, however, this `skin layer' has a principal meaning because it is a variable in the velocity expression.
For the lower values of the `skin layer' $ \sim \frac {1}{10} $ of the total thickness of the liquid layer, the thermocapillary flow dominates again. At lower values of the
`skin layer' just the convective flow becomes negligible.

\scriptsize

$\blacktriangle$

\normalsize
\subsection{Motion of a drop in viscous medium due to temperature gradient}

Let us consider now the problem of motion in a temperature field of a liquid drop suspended in another liquid.
Because of the temperature difference at different points of the drop surface, the later cannot remain at rest.
Instead, it will move from  the warmer regions with the lower surface tension toward the colder regions with the
higher surface tension.
In the same moment the moving drop will drag the surrounding liquid medium by applying some force to it.
There will also be an equal force in opposite direction
which the surrounding medium will apply onto the drop.
This reactive force will cause a drop movement in the
direction of the temperature gradient.

Let us estimate the order of the thermocapillary velocity of the drop. The characteristic force applied per unit length of the nonuniformly heated surface is of the order of
$ \left| \frac {d\sigma}{dT}\nabla T \right|$. This force is balanced
by the viscous forces$^{12)}$\blfootnote{$^{12)}$ $\mu$ and $\mu'$, $u$ and $u'$ are
the dynamic viscosities and liquid velocities inside a drop and in the surrounding liquid, respectively, $a$ is a drop radius (KM \& SB).}:
\begin{eqnarray}
\frac {\mu u}{a}+\frac {\mu' u'}{a}\sim \left| \frac {d\sigma}{dT}\nabla T \right|\,
\nonumber
\end{eqnarray}
Therefore we obtain
\begin{eqnarray}
u\sim \frac {a \left| \frac {d\sigma}{dT}\nabla T \right|}{\mu+\mu'}\,.
\nonumber
\end{eqnarray}

The exact value of the thermocapillary velocity can be found by solving the hydrodynamic equations
with the appropriate boundary conditions. We will use the hydrodynamic equations within the Stokes approximation, because in all practical cases both the thermocapillary velocity and the Reynolds number are small
\begin{eqnarray}
\nabla p=\mu \Delta \textbf{v}\,, & \,\,\,\,\,\nabla p'=\mu' \Delta \textbf{v}'\,, \nonumber\\
\textrm{div}\,\textbf{v}=0 \,, & \,\,\,\,\,\ \textrm{div}\, \textbf{v}'=0\,. \label{eq:13}
\end{eqnarray}
Here and below the variables with primes denote the liquid domain of the drop,
while the variables without primes relate to the surrounding liquid.
Let us consider the problem in the polar coordinate system attached
to the moving drop and with the origin in the drop center.
We chose the polar axes along
the temperature gradient. The velocity is defined to be positive when it is directed along the polar axis.
The boundary conditions far from the drop are
\begin{eqnarray}
v_{\tau}| _{r \rightarrow \infty}\rightarrow u \cos \theta\,,\,\,\,
v_{\theta}| _{r \rightarrow \infty}\rightarrow -u \sin \theta\,.
\label{eq:14}
\end{eqnarray}
At the drop surface the following components of the stress tensor are continuous$^{13)}$\blfootnote{$^{13)}$ The bulk and surface components of the stress tensor are shown here by upper indexes `$0$' and `$\sigma$',
respectively (KM \& SB).}
\begin{eqnarray}
 (p^0_{rr}+p_{rr}^{\sigma})|_{r=a}=p_{rr}^0\hspace{-3pt}'|_{r=a}\,,&&\nonumber\\
(p^0_{r\theta}+p_{r\theta}^{\sigma})|_{r=a}=p_{r\theta}^0\hspace{-3pt}'|_{r=a}\,.&&\label{eq:15}
\end{eqnarray}
In addition, at the drop surface the following conditions are true
\begin{equation}
v_r|_{r=a}=v'_r|_{r=a}=0\,,\,\,\,\,\,\, v_{\theta}|_{r=a}=v'_{\theta}|_{r=a}\,.
\label{eq:16}
\end{equation}
%
Here $v_r$, $v_{\theta}$, $v'_r$, $v'_{\theta}$ are the
corresponding components of the velocity field. The stress
tensor components are
\begin{eqnarray}
&{}&p^0_{rr}=-p+2\mu \frac {\partial v_r}{\partial r}\,,\nonumber\\
&{}&p^{\sigma}_{rr}=-\frac {2\sigma}{a}\,,\nonumber\\
&{}&p^0_{r\theta}=\mu \left( \frac {\partial v_{\theta}}{\partial r}+\frac {1}{r} \frac {\partial v_{r}}{\partial \theta}-\frac {v_{\theta}}{r} \right)\,, \label{eq:17}\\
&{}&p^{\sigma}_{r\theta}=\frac {1}{a}\frac {\partial \sigma}{\partial \theta}\,.\nonumber
\end{eqnarray}
Expressions for the variables with primes,
$p_{rr}^0\hspace{-3pt}'$, $p_{r\theta}^0\hspace{-3pt}'$,
are analogous to those for
$p_{rr}^0$ and $p_{r\theta}^0$ with the obvious substitutions of unprimed functions by their primed counterparts. Because of the uniaxial
symmetry of our problem, the velocity component $v_{\varphi}=0$
and all the variables of the problem are independent on the
azimuthal angle $\varphi$.
The surface tension is a part of boundary conditions and depends on the distribution of temperature on the drop interface.

We assume that the motion of the drop does not affect the distribution of temperature in the surrounding liquid, because the drop moves very slow. The temperature inside the drop is as if the drop had stayed at its particular location long enough. In addition, we assume the temperature gradient to be constant. Therefore the temperature distribution in both the drop and the surrounding liquid is described by the expression
\begin{equation}
T=T_c+|\nabla T|r\cos \theta\,,
\label{eq:18}
\end{equation}
where, $T_c$ is a temperature in the center of the drop. $T_c$ is a function of time
\begin{equation}
T_c=T_0+|\nabla T|ut\,.
\label{eq:19}
\end{equation}
Here $T_0$ is temperature in the center of the drop for initial drop position and $t$ is time. We also assume that the motion does not affect the drop spherical shape. Later, we will consider the feasibility of this assumption as well as its limits.

When the above assumptions are true, one can take the first spherical harmonic functions as a solution for the hydrodynamic equations [2].

The gradient of the surface tension caused by the inhomogeneous temperature on the interface is an origin of the motion on the drop surface. One can find the surface tension as a function of the angle $\theta$ using the following expression
\begin{eqnarray*}
\sigma=\sigma_{\pi/2}+\int_{\pi/2}^{\theta}\frac {d\sigma}{dT} \frac {\partial T}{\partial \theta}d \theta\,.
\end{eqnarray*}
As a first approximation, one can assume that the derivative ${d\sigma}/{dT}$ does not depend on temperature and therefore on angle $\theta$. In this case, taking into account Eq.~(18) one finds the surface tension
\begin{eqnarray}
\sigma=\sigma_{\pi/2}+ \frac {d\sigma}{dT} |\nabla T|a\cos \theta\,.
\label{eq:20}
\end{eqnarray}
%

Now we insert the last expression to the boundary conditions (16) getting seven equations to obtain the eight constants $a_0$, $b_0$, $a_1$, $b_1$, $a_2$, $b_2$, $b_3$, $u$ $^{14)}$\blfootnote{$^{14)}$ The following general form of the solution of Eqs.~(13) is used by Fedosov to express the velocity and pressure fields (KM \& SB):
\begin{eqnarray*}
&{}&v_r=\left(\frac {b_1}{r}+\frac {b_2}{r^3}+b_3\right)\cos \theta\,,\,\,\,\,\,\,\,\,\,\,\,\,v'_r=(a_1r^2+a_2)\cos \theta\,,\\
&{}&v_{\theta}=\left(-\frac {b_1}{2r}+\frac {b_2}{2r^3}-b_3\right)\sin \theta\,,\,\,\,\,\,v'_{\theta}=-(2a_1r^2+a_2)\sin\theta\,,\\
&{}&p=b_0+\mu \frac{b_1}{r^2}\cos \theta\,,\,\hspace{1.75cm}p'=a_0+10\mu'a_1r\cos \theta\,.
\end{eqnarray*}}.
As usual, one constant cannot be defined, so let us set $a_0=p_0$, where $p_0$ is a pressure in the center of the drop. The other constants are:
\begin{eqnarray}
&{}&b_0=p_0-\frac {2\sigma_{\pi/2}}{a}\,,\hspace{0.75cm}b_1=0\,,\nonumber\\
&{}&a_1=\frac 32 \frac {u}{a^2}\,,\,\hspace{1.7cm}b_2=-a^3u\,,\label{eq:21}\\
&{}&a_2=-\frac 32 u\,,\hspace{1.7cm} b_3=u=\frac 23 \frac {\frac {d\sigma}{dT} |\nabla T|a}{2\mu+3\mu'}\,.
\nonumber
\end{eqnarray}
The motion of liquid outside and inside the drop are described, correspondingly, by the following equations:
\begin{eqnarray*}
&{}&v_r=u\left(1-\frac{a^3}{r^3}\right)\cos \theta\,,\hspace{0.8cm}v'_r=\frac 32u\left(\frac{r^2}{a^2}-1\right)\cos \theta\,,\\
&{}&v_{\theta}=-u\left(1+\frac{a^3}{2r^3}\right)\sin \theta\,,\hspace{0.3cm}\,v'_{\theta}=\frac 32u\left(1-2\frac{r^2}{a^2}\right)\sin\theta\,,\\
&{}&p=p_0-\frac {2\sigma_{\pi/2}}{a}\,,\hspace{1.8cm}p'=p_0+15 \frac {\mu'}{a^2}ur\cos \theta\,.
\end{eqnarray*}
Velocity $u$ (this is velocity of the liquid at infinity) is opposite to the temperature
gradient because of the relation ${d\sigma}/{dT}<0$ . Velocity of the drop is $-u$, it is aligned along the temperature gradient. Thus, the drop moves from the cooler layers of liquid to the warmer ones. It is apparent that this motion promotes the temperature leveling-off.

Let us compare the thermocapillary velocity of the drop with a maximal velocity of the thermocapillary flow in the container
\begin{eqnarray*}
\left|\frac {u_{\mathrm{drop}}}{u_{\mathrm{container}}} \right|=\frac {8\mu a}{3(2\mu+3\mu')h}<\frac 43 \frac ah\,.
\end{eqnarray*}

Now, if one uses the characteristic values for the parameters
$\mu=0.01$~P, $\mu'=0.5$~P, $a=0.3$~mm, $|\nabla T|=0.1$~K/cm, $d\sigma/dT=-0.015$~erg/(cm$^2 \cdot$K)
in the drop velocity expression,
then one can estimate the drop velocity  $u=2\cdot10^{-4}$ cm/s$^{15)}$\blfootnote{
$^{15)}$ In fact, $|\nabla T|$ can be $\sim 1$ K/cm (KM \& SB).}.

We have mentioned above that assumption of the constant spherical shape of the drop should constrain our theory.
Let us consider this constraint now. We believe that the drop keeps its shape nearly spherical when the surface tension on the drop interface is nearly constant, i.e., $|\sigma -\sigma_{\pi/2}| \ll \sigma_{\pi/2}$, where $\sigma_{\pi/2}$ is the surface tension on the drop's equator and $\sigma$ is the surface tension in the any other point of the drop interface. Assuming this we can write
\vspace{-0.2cm}
\begin{equation}
({d\sigma}/{dT}) |\nabla T|a/{\sigma_{\pi/2}} \ll 1\,.\label{eq:22}
\end{equation}
To make clear in which parameter regions our theory is applicable, let us consider a numerical example. The equator's surface tension depends on the equator's temperature, however, it
is not significant regarding our estimate of the theory's validity. Therefore for the values of the parameters given above we have
$|\nabla T|a \ll 50\, \mathrm{K}\,.$
This inequality requires the temperature change along the drop perimeter to be less than $50$ K. It is clear that this inequality is almost always true.  Indeed, for $|\nabla T|=0.1$~K/cm we should require the following inequality $a \ll 5$ m to be true. It is obvious that all drops in real experiments satisfy this condition.

One can show that the thermocapillary drop velocity can overcome (and sometimes, for some cases does so significantly) the electrocapillary drop velocity for weakly conductive drops.

I would like to thank V.G. Levich for giving this problem to my consideration and for his constant advice during my work.

\vspace{-0.4cm}
\subsection{CONCLUSIONS}
\vspace{-0.3cm}

1. Two cases of the flow of liquid caused by the temperature gradient in the interfacial layer (thermocapillary flow) are considered.
2. Thermocapillary flow can overcome convective flow when the interfacial heating layer is reasonably small.
3. The limits of the theory's applicability are discussed.

\newpage

\begin{widetext}

\vspace{-0.4cm}

\section{Biographical sketch}
\vspace{-0.6cm}
\begin{figure}[h]
\includegraphics[width=6.5cm]{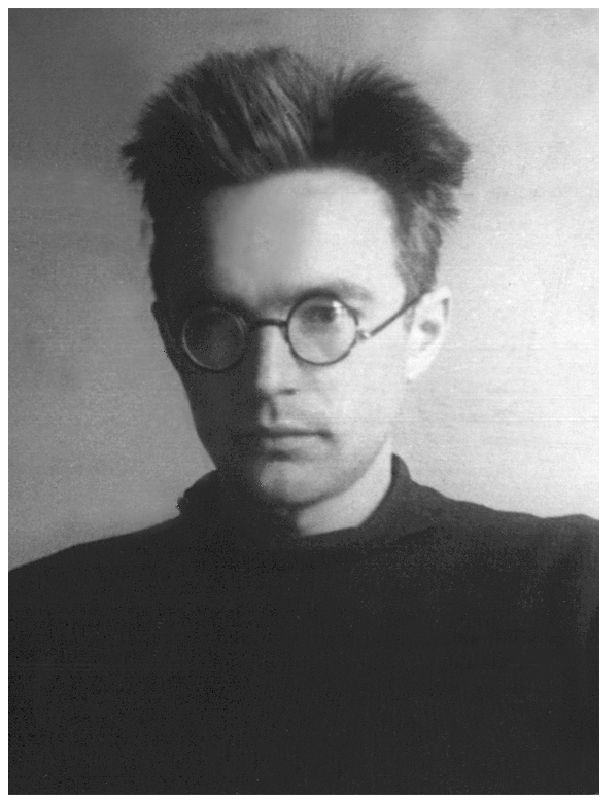}
\end{figure}
\end{widetext}
\noindent
Alexander I. Fedosov (13.09.1923--05.10.1999) was born in a small village in the northern region of the Saratov province of the former Soviet Union. After graduating from school with a gold medal, he became a student of the Moscow Pedagogical Institute. He graduated in 1945, and continued to the postgraduate program under the supervision of V.G. Levich. In 1948, A.I. Fedosov successfully defended his PhD thesis entitled `Some problems of the theory of surface phenomena' in the Moscow Pedagogical Institute. He was unable of permanently stay in Moscow after getting his PhD, and he moved to Chita (Trans-Baikal region, Siberia) - a city 6,500 km away from Moscow. In the following period, from 1949 to 1955, A.I. Fedosov worked in the Chita Pedagogical Institute as a senior lecturer of physics. In 1955, A.I Fedosov moved to Kuibyshev (nowadays known as Samara), where he worked as an associate professor and a Head of the Department of Physics first in the Pedagogical Institute, then in the Agricultural Institute and at the end of his career in the Aviation Institute.

\vspace{0.8cm}

\subsection{List of main publications of Fedosov$^{16)}$} \blfootnote{\\  $^{16)}$ For some reasons, unknown for us, it was only in 1955-1956 when
A.I.~Fedosov had published the three parts of his PhD thesis
as three independent papers  (b), (c), and (d). We point out that these papers are almost textually identical to the corresponding chapters of his PhD thesis (KM \& SB).}

\vspace{-0.1cm}

(a) A.I.~Fedosov, Ph.D. thesis. \emph{Some problems of the theory of surface phenomena.}  Moscow State Pedagogical Institute. 1948. (In Russian).

(b) A.I.~Fedosov, \emph{Electrocapillary motion of drops}, Zhurnal Fizicheskoi Khimii \textbf{29}, N5, p. 822-831 (1955). (In Russian).

(c) A.I.~Fedosov, \emph{Effect of surfactant on the motion of drops in liquids}, Zhurnal Fizicheskoi Khimii \textbf{30}, N1, p. 223-227 (1956). (In Russian).

(d) A.I.~Fedosov, \emph{Thermocapillary motion}, Zhurnal Fizicheskoi Khimii \textbf{30}, N2, p. 366-373 (1956). (In Russian).

(e) A.I.~Fedosov, \emph{Drag of bubble movement by surfactant at moderate Reynolds numbers}, Zhurnal Fizicheskoi Khimii \textbf{33}, N8, p. 1681-1686 (1959). (In Russian).


\vspace{-0.3cm}


\begin{thebibliography}{10}

\vspace{-0.3cm}
\bibitem{1}
\emph{The current state of hydroaerodynamics of viscous fluid, V. 2.} Ed.: S. Goldshtein,
Moscow, 1948. (In Russian).

\bibitem{2}
H. Lamb, \emph{Hydrodynamics} (Dover, New York, 1945).

\vspace{0.4cm}

{\bf References and footnoted supplements added by KM and VB}
\vspace{0.3cm}

\bibitem{3}
A.I. Fedosov, Ph.D. thesis. \emph{Some problems of the theory of surface phenomena.}  Moscow State Pedagogical Institute. 1948.
(In Russian).

\bibitem{4}
V.G. Levich, \emph{Physicochemical hydrodynamics.} Englewood Cliffs, N.J., Prentice-Hall. 1962.

\vspace{0.3cm}
\bibitem{5}
R.V. Birikh, \emph{Thermocapillary convection in a horizontal
layer of liquid}, Journal of Applied Mechanics and Technical Physics \textbf{7}, N 3, pp. 43-44 (1966); DOI: 10.1007/BF00914697.


\end{thebibliography}
\end{document}